\documentclass[12pt]{article}

\usepackage{amssymb}

\begin{document}

\begin{titlepage}

\begin{flushright}
\end{flushright}
\vskip 2.5cm

\begin{center}
{\Large \bf Testing Photons' Bose-Einstein Statistics\\
With Compton Scattering}
\end{center}

\vspace{1ex}

\begin{center}
{\large Brett Altschul\footnote{{\tt baltschu@physics.sc.edu}}}

\vspace{5mm}
{\sl Department of Physics and Astronomy} \\
{\sl University of South Carolina} \\
{\sl Columbia, SC 29208} \\
\end{center}

\vspace{2.5ex}

\medskip

\centerline {\bf Abstract}

\bigskip

It is an empirical question whether photons always obey Bose-Einstein statistics, but
devising and interpreting experimental tests of photon statistics can be a challenge.
The nonrelativistic cross section for Compton scattering illustrates how a small
admixture $\nu$ of wrong-sign statistics leads to a loss of gauge invariance; there
is a large anomalous amplitude for scattering timelike photons. Nevertheless, one can
interpret the observed transparency of the solar wind plasma at low frequencies as a
bound $\nu<10^{-25}$ if Lorentz symmetry is required. If there is instead a
universal preferred frame, the bound is $\nu<10^{-14}$, still strong compared with
previous results.

\bigskip

\end{titlepage}

\newpage

The connection between spin and statistics
is one of the most
profound results in quantum physics. In a relativistic quantum theory, imposing
a small number of reasonable conditions (including Lorentz invariance and stability
of the vacuum state) necessitates that particles with integer spin have wave
functions symmetric under exchange and those with half integer spin have antisymmetric
wave functions. This is a theorem in axiomatic field theory, but whether it is really
true of physical particles is a separate question. If any deviation from the usual
spin-statistics relationship were uncovered in the laboratory, this would be a
discovery of profound importance; it would be a sure sign of genuinely new physics,
and it would necessitate a violation of some other basic axiom that has been
believed to hold.

Recently, a remarkable
experiment using a two-photon atomic transition has been interpreted as a
new test of photon statistics~\cite{ref-english}. The transition rate is
affected by the symmetry state of the two photons involved. However, the strategy
used in this laboratory test can
actually be applied to a much broader class of experiments, and
it is important that a violation of Bose-Einstein (BE) statistics would not be
a purely quantum mechanical effect. Classical electrodynamics already
contains many of the correlations that characterize BE
behavior. It is therefore possible to test the statistics of photons by
looking only at classical (i.e. tree level) processes.

Studies of spin-statistics violations have historically focused more on small
deviations from Fermi-Dirac (FD) statistics. There are two good reasons for
this. First, it is quite easy to conceive of an experiment to test the
Pauli Exclusion Principle: one simply tests whether it is genuinely
impossible for two electrons to occupy the same quantum state, something
that can be determined from atomic spectroscopy~\cite{ref-bartalucci}.
Second, the
classical limit of a fermionic field means having a single quantum
present, while the classical limit of a bosonic field involves a large
number of quanta. But while it is clearly meaningful to have a bosonic
field with only one quanta present, one cannot have a fermionic field with
many. This means that a small change in the statistics of a
spin-$\frac{1}{2}$ field need not have large, obvious
consequences in the classical limit; it is easier to envision a form of
new physics that occasionally allows more than one particle in a given
state than a new physical principle that occasionally reduces the allowed
number from infinity to something finite. This makes testing photons statistics an
interesting challenge.

The experiment in~\cite{ref-english}
used two laser beams, aligned nearly parallel. The two beams'
frequencies were swept up and down, keeping their sum constant and in resonance with
a $\Delta J=1$ atomic transition. Most of the time, the photons could combine
(with one from
each beam) to excite this transition. However, the requirement that two photons be in
a symmetric state dictates that two photons with equal momenta cannot be in a state
with total angular momentum $J=1$. Therefore, if the two laser frequencies were
equal, the atomic transition rate would vanish. As the frequencies approached
equality, BE correlations ensured the transition rate went smoothly to
zero. The object of study was these correlations; the experiment did not look
directly at the question of whether it is possible for multiple photons to be present
in the same quantum state.

If the BE correlations were not present exactly as expected, one would
expect there to be a small but nonzero transition rate even if photons from the two
laser beams had precisely equal momenta. Such a modification to the photons'
statistics was modeled as follows. In the transition rate, there are two
diagrams, in which the photons are absorbed in opposite orders. Instead of simply
calculating the sum of the two diagrams, there could be a small admixture of
the difference of the diagrams. This difference would normally be
taken if photons were fermions, so this model means having a small admixture of
wrong-sign statistics. In addition to the usual transition rate $w_{BE}$
(calculated with BE statistics), there is a small fraction $\nu$ of the
FD transition rate $w_{FD}$, making the total rate $w=w_{BE}+\nu w_{FD}$.

Note that the test described in~\cite{ref-english} does not involve electromagnetic
excitations that can be treated as non-bosonic over their entire time of
existence. Instead, the electromagnetic field is prepared in a
coherent and intrinsically bosonic laser state. In models in which there exists a
small admixture of FD photons among the usual BE ones, the use of
lasers may be problematic. The techniques used to enhance the intensity of a laser
beam may increase the number of bosonic photons present, but the number of
fermionic photons will remain zero or one. Thus, increasing the laser power may
enhance only the absorption rate for conventional photons, and the
FD signal may be swamped by BE noise.

For this reason, the two-photon absorption measurement is most sensitive to the
possibility of photons that behave, for the vast majorities of the lifetimes, as
bosons; however, they briefly behave according to FD statistics in the time
period immediately surrounding the atomic transition. This
restricts the applicability of
these bounds to only certain types of spin-statistics violations. This
restriction will
apply to any experimental technique that uses large electromagnetic field amplitudes.

Compton scattering, viewed as a scattering process with one fermion and one photon,
need not suffer from the same limitation. The photon could be entirely fermionic,
over the entire course of its existence. This makes Compton scattering (and other
scattering processes that do not involve coherent assemblages of photons) sensitive
to more possible forms of spin-statistics violation than tests with lasers or
macroscopic fields. However, the modified Compton scattering cross section can also
be applied to situations with coherent fields, such as the scattering of radio waves
in diffuse plasmas. In that case, a study using Compton scattering would be sensitive
to the same kinds of statistical correlations as the two-photon absorption
experiment.

No complete theory in which photons do not obey BE statistics
is known. We shall therefore follow the
interpretational methodology of~\cite{ref-english}; we
consider a single process, which receives contributions from two distinct
photon interaction diagrams. The statistical properties of the
electromagnetic excitations involved determine the relative sign between
the two diagrams. Depending on the statistics, the amplitudes may
interfere constructively or destructively.
We shall show, by example, that inverting the relative sign between the two diagrams
that can contribute to an ${\cal O}(e^{2})$ interaction amplitude destroys a number
of key properties of quantum electrodynamics. Gauge invariance is lost, as may be
Lorentz invariance.
Nevertheless, we can estimate how strongly the
violation of BE statistics for photons is constrained by experiment.

The matrix element squared for Compton scattering (summed and averaged over fermion
polarizations, and using natural units in which $\hbar=c=\varepsilon_{0}=1$) is
\begin{eqnarray}
\frac{1}{2}\sum_{f\,{\rm spins}}|{\cal M}|^{2} & = & 
\frac{e^{4}}{2}
(\epsilon'^{*}_{\mu}\epsilon'_{\rho})(\epsilon_{\nu}\epsilon^{*}_{\sigma})\,{\rm tr}
\left[(\!\not\!p'+m)\left(\frac{\gamma^{\mu}\!\not\!k\gamma^{\nu}+2\gamma^{\mu}
p^{\nu}}{2p\cdot k}\pm\frac{\gamma^{\nu}\!\not\!k'\gamma^{\mu}-2\gamma^{\nu}
p^{\mu}}{2p\cdot k'}\right)\right. \nonumber\\
& & \left.\times(\!\not\!p+m)
\left(\frac{\gamma^{\sigma}\!\not\!k\gamma^{\rho}+2\gamma^{\rho}
p^{\sigma}}{2p\cdot k}\pm\frac{\gamma^{\rho}\!\not\!k'\gamma^{\sigma}-
2\gamma^{\sigma}p^{\rho}}{2p\cdot k'}\right)\right].
\end{eqnarray}
$p$ and $p'$ are the initial and final fermion momenta; $k$ and $k'$ are the
corresponding photon momenta; $\epsilon(k)$ and $\epsilon'(k')$ are the incoming
and outgoing polarization vectors. The $+$ signs
between the terms are the usual ones; the $-$ signs
correspond to the possibility of inverted statistics. There are two Feynman diagrams
that contribute to this process; they differ in the order of the photon emission and
absorption vertices along the fermion line. (In nonrelativistic scattering with the
wrong photon statistics, this is
the same as the time order of the two photon interaction events.) Normally the
two diagrams are summed, in accordance with BE statistics.

We shall concentrate on the low-energy regime, in
which the fermion is initially at rest and $\omega\ll m$. Energy-momentum
conservation then dictates that $\omega'\approx\omega$ and
$p'\approx p=(m,\vec{0}\,)$. The largest contribution to the scattering cross section
then comes from the electron propagator terms with $p$ in the numerator. Since
$p\cdot k'\approx p\cdot k=m\omega$, this becomes
\begin{eqnarray}
\frac{1}{2}\sum_{f\,{\rm spins}}|{\cal M}|^{2} & = & 
\frac{e^{4}}{2}
(\epsilon'^{*}_{\mu}\epsilon'_{\rho})(\epsilon_{\nu}\epsilon^{*}_{\sigma})\,{\rm tr}
\left\{(m\gamma^{0}+m)\left[\frac{2\gamma^{\mu}(mg^{0\nu})}{2m\omega}\pm
\frac{-2\gamma^{\nu}(mg^{0\mu})}{2m\omega}\right]\right. \nonumber\\
& & \left.\times
(m\gamma^{0}+m)\left[\frac{2\gamma^{\rho}(mg^{0\sigma})}{2m\omega}\pm
\frac{-2\gamma^{\sigma}(mg^{0\rho})}{2m\omega}\right]\right\} \\
& = & (\epsilon'^{*}_{\mu}\epsilon'_{\rho})(\epsilon_{\nu}\epsilon^{*}_{\sigma})
\frac{e^{4}m^{2}}{2\omega^{2}}\,{\rm tr}[(\gamma^{0}+1)\gamma^{\mu}(\gamma^{0}+1)
\gamma^{\rho}g^{0\nu}g^{0\sigma} \nonumber\\
& & \mp(\mu\leftrightarrow\nu)\mp(\rho\leftrightarrow\sigma)
+(\mu\leftrightarrow\nu,\rho\leftrightarrow\sigma)].
\end{eqnarray}
Since ${\rm tr}[(\gamma^{0}+1)\gamma^{\mu}(\gamma^{0}+1)\gamma^{\rho}]=
8g^{0\mu}g^{0\rho}$, the matrix element squared is
\begin{equation}
\label{eq-Mtimelike}
\frac{1}{2}\sum_{f\,{\rm spins}}|{\cal M}|^{2}=
(\epsilon'^{*}_{\mu}\epsilon'_{\rho})(\epsilon_{\nu}\epsilon^{*}_{\sigma})
\frac{16e^{4}m^{2}}{\omega^{2}}g^{0\mu}g^{0\nu}g^{0\rho}g^{0\sigma}
\end{equation}
for wrong-sign
statistics, while it vanishes (at this order in
$\frac{m}{\omega}$) for BE statistics.
The total cross section---including both the usual term and a small admixture
$\nu$ of wrong-sign statistics---is
$\sigma=\sigma_{T}\left(1+6\frac{m^{2}}{\omega^{2}}\Sigma\nu\right)$,
where $\sigma_{T}=\frac{8\pi\alpha^{2}}{3m^{2}}$ is the Thomson cross
section, and
$\Sigma=|\epsilon'_{0}|^{2}|\epsilon_{0}|^{2}$ represents the photon
polarization expression, which must be properly interpreted. The result remains
unchanged if the scatterer is a composite particle with form factors $F_{1}$ and
$F_{2}$.

The matrix element for the wrong statistics case has many unusual features.
The scattering is isotropic, in contrast to the usual case, and
the cross section is large---${\cal O}(\frac{m^{2}}{\omega^{2}})$.
The contributions that make it large come from virtual
intermediate states that are nearly on shell; they have energy defects of approximately
$\pm\omega$ and are very close to being real states. Normally, there is
destructive interference between the two diagrams, wiping out those terms
in the matrix element that diverge as $\omega\rightarrow0$. This interference
is ensured by the BE statistics of the photons; the incident
and scattered photons must be in a symmetric state.  (Note that, depending
on the temporal order in which absorption and emission occur, these two
photons may or may not be physically present at the same instant in time;
however, the statistical relationship between the two is required whether
or not they actually exist simultaneously.) Without the conventional symmetry
requirement, the cancellation is not complete, and large new terms appear
in the cross section. These terms modify the Thomson result, which is ordinarily
entirely classical in character; however, since the large modifications
are tied to the presence of nearly on shell virtual states, there is
probably no way to formulate the wrong-statistics photon scattering problem
using only classical techniques.

Moreover, in the wrong-statistics version, gauge invariance is obviously
violated; the amplitude is not transverse to the
photon momenta: $k_{\nu}{\cal M}^{\mu\nu}\neq0$. The cancellation between the two
diagrams is crucial to maintaining this Ward identity in the conventional case. The
modified behavior represents a large deviation from standard physics,
and the classical interaction
Lagrangian for interactions between the charged particle and the magnetic field is
no longer the usual $-ev^{\mu}A_{\mu}$. Gauge invariance is lost if FD statistics are
used in the two-photon absorption experiment as well. The loss of gauge symmetry
could have even
more drastic consequences in other processes; if the wrong-statistics amplitude were
included in calculations of the vacuum polarization, that diagram would generate a
divergent photon mass.

The $|{\cal M}|^{2}$ appearing in
(\ref{eq-Mtimelike}) corresponds to scattering exclusively to and from timelike
photon states; those are ghost states with negative norm. Normally, the contributions
from timelike states are precisely canceled by contributions from longitudinal
polarization states; this is a consequence of gauge invariance, as expressed by the
Ward identity. A sum over polarization vectors
$\sum\epsilon_{\nu}\epsilon^{*}_{\sigma}$ may be replaced by $-g_{\nu\sigma}$, which
produces a manifestly Lorentz invariant expression.

However, the failure of the Ward identity when the wrong statistics are used makes
interpreting (\ref{eq-Mtimelike}) a nontrivial matter. If gauge symmetry is broken,
there must be an ancillary condition on the electromagnetic field (e.g. the Lorenz
condition $\partial^{\mu}A_{\mu}=0$ in massive electrodynamics), but it is not clear
what the condition should be in the theory of wrong-statistics photons.
The key question is what
external photon polarization vectors $\epsilon$ and $\epsilon'$
are appropriate for determining the physical
cross section.
There are two particularly obvious ways of choosing these vectors,
which differ in their physical consequences. The two possibilities can essentially
be characterized by how many external spin states are to be summed over, but it is
not clear which (if any) of the interpretations can be assigned completely
consistent interpretations.

The first possible interpretation is simply to retain the prescription
$\sum\epsilon_{\nu}\epsilon_{\sigma}^{*}\rightarrow-g_{\nu\sigma}$ in the
spin-summed cross section. This has the advantage of preserving Lorentz invariance,
by summing over polarization vectors in all four possible spacetime directions.
This amounts to setting $\Sigma=\frac{1}{2}$ in the spin-averaged cross-section.


The second
possibility is to break Lorentz symmetry, which makes it possible
to choose
two purely spacelike vectors as the only physical external photon polarizations. Breaking Lorentz symmetry might seem like a drastic and unnecessary additional
modification of the physics, but in many ways the theory
with broken Lorentz symmetry seems to be better behaved than the theory utilizing
the Lorentz-invariant polarization prescription described above. The
Lorentz-symmetric option invokes additional polarization states, which generally
interact with charged matter as strongly (or more strongly) than the usual transverse
polarizations. Yet there are only two real electromagnetic field states that come
into black body equilibrium on observable time scales.
On the other hand, in the absence of Lorentz invariance, these extra states can be
excluded from the theory. Since it is only the timelike photon states that appear in
the scattering cross section (\ref{eq-Mtimelike}), having only purely spacelike
polarizations would almost seem to alleviate the problem with (\ref{eq-Mtimelike})
entirely. Yet using just two purely spacelike external polarization vectors does not
actually address all the problems. There remain three significant challenges for the
Lorentz-violating version of this theory.

The first challenge comes from the fact that (\ref{eq-Mtimelike}) is only an
approximation. There are other novel terms in the wrong-statistics cross section that
do affect the scattering of purely spacelike photons; these terms are
smaller, lacking
the large enhancement factor $\frac{m^{2}}{\omega^{2}}$, but they are still 
comparable in size to the usual Thomson cross section and thus are potentially
observable, even
for small values of $\nu$. The second issue derives from the fact this it is, of
course, only possible to choose $\epsilon$ and $\epsilon'$ as purely spacelike in
one particular frame, which would be a preferred frame for the universe. The only
likely candidate for such a frame is the rest frame of the cosmic microwave
background (CMB), and the solar system is moving relative to the CMB at a speed
$v_{CMB}\approx2\times10^{-3}$. If the time components of $\epsilon$ and $\epsilon'$
vanish in the CMB frame, the time components will be $\sim v_{CMB}$ in a terrestrial
laboratory frame. So the cross section for scattering off a fermion that is
stationary in the lab frame will receive a $\Sigma\sim v_{CMB}^{4}$  contribution
from (\ref{eq-Mtimelike}). This is a suppression of roughly $10^{-11}$, but it does
not rule out observable consequences.

The third challenge for the Lorentz-violating scenario is that it must confront the
bounds that have already been placed on spin-independent Lorentz symmetry violations
in electron-photon interactions. The best tests of boost invariance in such processes
use astrophysical synchrotron and inverse Compton data to place bounds at the
$10^{-14}$--$10^{-19}$ levels~\cite{ref-altschul7}. However, the reliability of such
bounds is questionable if the electron-photon interaction differs from the usual one;
the usual inferences about what processes are responsible for the observed spectra
of given sources may not be valid. The best laboratory bounds are based on
observations of similar processes at the Large Electron-Positron Collider (LEP).
Those bounds should be considered more secure, since the processes involved are known
with relative certainty. The best LEP bounds on the same quantities are at the
$10^{-12}$--$10^{-15}$ levels~\cite{ref-hohensee1,ref-altschul22}.
It is not precisely clear how
these bounds apply to the Lorentz-violating theory of wrong-statistics photons, but
some effect would be expected.

Despite these challenges, it is still possible to estimate how much FD statistics
can be present in Compton scattering processes, by looking at the scattering of
low-energy photons. For this purpose, it is advantageous to look at the long-distance
propagation of radio waves through plasmas. (Diffuse plasmas are preferable, since
collective behavior takes over at frequencies below
$\omega_{p}=\sqrt{n_{e}e^{2}/m}$, where $n_{e}$ is the electron density.)
In order to have a substantial amount of
anomalous, frequency-dependent scattering, the optical depth due to the
wrong-statistics process, $6\frac{m^{2}}{\omega^{2}}(nL\sigma_{T})\Sigma\nu$
(where $n$ is the density of scatterers, so $nL\sigma_{T}$ is the depth due to
Thomson scattering) must be $\sim1$ or larger.

The Wind satellite observed radio emissions down to frequencies
$\frac{\omega}{2\pi}\approx70$ kHz coming from Earth (over a distance
$L>1.3\times10^{11}$ cm) with no unusual scattering by the solar wind
evident~\cite{ref-kaiser}.
Since the anomalous scattering is stronger for heavier particles, the relevant
density $n$
is that of protons in the solar wind, which was $n_{p}\approx 10$--20 cm$^{-3}$ along
the radiation's path. The transparency implies the bound
\begin{equation}
\Sigma\nu<\frac{\omega^{2}}{6m_{p}^{2}n_{p}L\sigma_{T}}<1.8\times10^{-26}.
\end{equation}
For the version of the theory with broken Lorentz symmetry, in which the CMB frame
is the preferred frame, this implies a conservative bound on $\nu$ of
\begin{equation}
\nu<10^{-14}.
\end{equation}
The Lorentz-invariant theory, with strongly coupled longitudinal and timelike
photons, is harder to interpret, but the numerical bounds on $\nu$ are stronger,
\begin{equation}
\nu<10^{-25}.
\end{equation}
Each of these bounds is much stronger than the $\nu<4\times10^{-11}$ result
from~\cite{ref-english}.


Of course,
the two-photon absorption and Compton scattering data do not measure precisely
the same thing. In each case, the rate for a process is calculated by conventional
means, except that a possible change in the photons statistics is inserted at one
point (with the remainder of the calculation tacitly assuming BE behavior).
In the absence of any underlying theory to motivate the change in
statistics, it is tricky to relate the results of significantly different experiments
(although if there is violation of BE statistics in one case, it would
be natural to expect a similar violation in other experiments). So the two-photon
absorption and Compton scattering bounds should be seen as complementary, rather than
directly competitive.

Moreover, simple experiments such as these could also be interpreted of tests of
other basic principles; and again, the two experimental scenarios would not generally
be sensitive to exactly the same things. For example, the two-photon transition
experiment
could be interpreted as a test not of photon statistics but of whether photons states
always have spin angular momentum $J=1$. In Lorentz-violating or nonlinear theories
of electrodynamics, this fact is not assured. Unusual behavior in Compton scattering
could also be indicative of a different form of Lorentz violation; even though the
Thomson cross section can be derived from a very limited set of
assumptions (see e.g.~\cite{ref-thirring}),
${\cal O}(\omega^{-2})$ modifications are
possible if boost symmetry is broken~\cite{ref-altschul2}.

\end{document}